\documentstyle[preprint,aps]{revtex}
\begin{document}
\title{Evidences for Tsallis non-extensivity on CMR manganites}
\author{M.S. Reis$^{a\ast }$, J.C.C.Freitas$^{b}$, M.T.D.Orlando$^{b}$, E.K. Lenzi$%
^{a}$ and I.S.Oliveira$^{a}$}
\address{(a) Centro Brasileiro de Pesquisas F\'{i}sicas \\
Rua Dr. Xavier Sigaud 150, Rio de Janeiro-RJ 22290-180, Brazil\\
(b) Departamento de F\'{i}sica, Universidade Federal do\\
Espirito Santo,\ Vit\'{o}ria-ES, 2906-900, Brazil.\\
\qquad \\
{\small *Corresponding author (present address):}\\
{\small Departamento de F\'{i}sica, Universidade de Aveiro,}\\
{\small 3810 Aveiro, Portugal.}\\
{\small e-mail: marior@cbpf.br}\\
{\small Fax.: +351 234 424 965}\\
{\small Tel.: \ +351 234 370 356}}
\maketitle

\begin{abstract}
We found, from the analysis of $M$ vs. $T$ curves of some manganese oxides
(manganites), that these systems do not follow the traditional
Maxwell-Boltzmann statistics, but the Tsallis statistics, within the {\em %
normalized} formalism. Curves were calculated within the mean field
approximation, for various ferromagnetic samples and the results were
compared to measurements of our own and to various other authors published
data, chosen at random from the literature. The agreement between the
experimental data and calculated $M_{q}$ vs. $T^{\ast }$ curve, where $%
T^{\ast }$ is an effective temperature, is excellent for all the compounds.
The entropic parameter, $q$, correlates in a simple way with the
experimental value of $T_{c}$, irrespect the chemical composition of the
compounds, heat treatment or other details on sample preparation. Examples
include $q<1$ (superextensivity), $q=1$ (extensivity) and $q>1$
(subextensivity) cases.
\end{abstract}

Manganese oxides, or simply manganites, have, to a great extent, dominated
the literature on magnetism for the last five years \cite{ref1}. The number
of yearly published papers on the subject since 1993 to present date,
amounts to over 2000. All this interest lies on at least three different
reasons: (i) the rich phase diagram of manganites exhibits a variety of
transport, structural and magnetic phenomena\cite{ref1,ref2}, which
stimulates new models in condensed matter\cite{ref1a}-\cite{ref1l} ; (ii)
manganites can present the so-called {\em colossal magnetoresistance }
(CMR), and therefore are interesting systems for industrial applications 
\cite{ref3} and, (iii) samples are relatively easy to prepare \cite{ref1}.

In the literature of manganites, various models have appeared in different
attempts to reproduce the electric and magnetic properties of these systems.
Krivoruchko et al \cite{ref1a}, Nunez-Regueiro et al. \cite{ref1b} and
Dionne \cite{ref1c} are interesting examples of multiparameter models, but
which failed to achieve full agreement to experimental data. Ravindranath et
al. \cite{ref1d} compare resistivity data in La$_{0.6}$Y$_{0.1}$Ca$_{0.3}$MnO%
$_3$ to different two-parameter models which do not agree to each other in
the low-temperature range. Other interesting attempts can be found in Rivas
et al. \cite{ref1e}, Hueso et al. \cite{ref1f}, Heremano et al. \cite{ref1g}%
, Pal et al. \cite{ref1h}, Philip et al. \cite{ref1i}, Szewczyk et al. \cite
{ref1j}, Viret et al. \cite{ref1k} and Tkachuk et al. \cite{ref1l}. None of
these obtained plain agreement between experiment and theory, irrespect
their number of adjusting parameters and approach.

Another rather different area which has been growing at an analogous rate is
the Tsallis generalized statistics \cite{ref20}, and its applications \cite
{ref5,ref10}. This is based on the definition of generalized entropy \cite
{ref20,ref5}:

\begin{equation}
S_{q}=k_{B}\frac{1-\sum_{i}p_{i}^{q}}{q-1}  \label{eq1}
\end{equation}
where $q$ is the {\em entropic index} and $p_{i}$ are probabilities
satisfying $\sum_{i}p_{i}=1$. The above formula converges to the usual
Maxwell-Boltzmann definition of entropy in the limit $q\rightarrow 1$ \cite
{ref20,ref5}.

Applications of Eq. (\ref{eq1}) to condensed matter include: Ising
ferromagnets \cite{ref6}, molecular field approximation \cite{ref7},
percolation problems \cite{ref8}, Landau diamagnetism \cite{ref16},
electron-phonon systems and tight-binding-like Hamiltonians \cite{ref17},
metallic \cite{ref9} and superconductor \cite{ref23} systems, etc.

Maximization of Eq. ($\ref{eq1}$) subjected to the normalized $q$%
-expectation value of the Hamiltonian \cite{ref9a}: 
\begin{equation}
U_{q}=\frac{Tr\{{\cal H}\rho ^{q}\}}{Tr\{\rho ^{q}\}}  \label{eq2}
\end{equation}
and the usual normalization of the density matrix $Tr\{\rho \}=1$, yields
the following expression for the density matrix $\rho $: 
\begin{equation}
\rho =\frac{1}{Z_{q}}[1+(1-q)\tilde{\beta}({\cal H}-U_{q})]^{1/(1-q)}
\label{eq3}
\end{equation}
where $Z_{q}={\mbox {Tr}}[1+(1-q)\tilde{\beta}({\cal H}-U_{q})]^{1/(1-q)}$
is the partition function, $\tilde{\beta}=\beta /c_{q}$ and $c_{q}=Tr\{\rho
^{q}\}$. The magnetization of a specimen is, accordingly, given by: 
\begin{equation}
M_{q}=\frac{Tr\{\mu \rho ^{q}\}}{Tr\{\rho ^{q}\}}  \label{eq2a}
\end{equation}
Note that Eqs.(\ref{eq2}) and (\ref{eq2a}) need to be solved
self-consistently, since the density matrix $\rho $\ [Eq.($\ref{eq3}$)]
depends on the hamiltonian ${\cal H}=-\mu \lbrack B_{0}+\lambda M_{q}]$\ and
its q-expectation value, U$_{q}$. In the hamiltonian above, $B_{0}$\
represents the external magnetic field, and $\lambda $\ the molecular field
parameter. Similar calculation in the context of many particle system
employing the normalized approach has been done for the quantum statistics 
\cite{ref17i,ref17a}.

Eq.(\ref{eq3}) can be written in a more convenient form in terms of $\beta
^{\ast }$, defined as: $\beta ^{\ast }\equiv \tilde{\beta}/(1+(1-q)\tilde{%
\beta}U_{q})$ \cite{ref9a}. In particular, to analyze the physical system
described here, the quantity $1/k_{B}\beta ^{\ast }\equiv T^{\ast }$
represents a temperature scale against which this quantity $M_{q}$ will be
compared to experimental results. A discussion about the concept of
temperature and Lagrange parameters in Tsallis statistics can be found in\
literature \cite{ref9a,ref17i,ref17l,ref17m}.

In this Letter we argue, based on published experimental results, that
manganites are non-extensive objects. This property appears in systems where
long-range interactions and/or fractality exist, and such features have been
invoked in recent models of manganites, as well as in the interpretation of
experimental results. In a recent review \cite{ref17b}, Dagotto and
co-workers emphasize the role of the competition between different phases to
the physical properties of these materials. Various authors have considered
the formation of micro-clusters of competing phases, with {\em fractal}
shapes, randomly distributed in the material \cite{ref17c,ref17d}, and the
role of {\em long-range interactions} to phase segregation \cite
{ref17e,ref17f}. Important experimental results in this direction have also
been reported by Marethew {\it et al.} \cite{ref17g}, and Fiebig {\it et al.}
\cite{ref17h}. Particularly insightful is the recent paper of Satou and
Yamanada \cite{ref17j} who derived a {\em Cantor} spectra for the
double-exchange hamiltonian, basis of theoretical models of manganites. In
spite of these evidences, Tsallis statistics, to the best of our knowledge,
has never been used in the context of manganites.

Bulk samples of La$_{0.89}$Sr$_{0.11}$Mn$_{1-x}$Cu$_{x}$O$_{3}$ (x = 0;
0.07) were produced by standard solid-state reaction of high-purity La$_2$O$%
_3$, SrCO$_3$, MnO$_2$ and CuO powders mixed in stoichiometric proportions.
The mixed powder was pressed into pellets and heat-treated at 930 $^o$C in
atmospheric air during four days with three crushing/pressing procedures
intermediating the treatment. Then, the pellets were heat-treated at 1350 $%
^o $C under oxygen flow during 48 h, with one intermediate crushing and
pressing. X-ray diffraction (XRD) indicated the formation of single-phase
samples (rhombohedral cell). Scanning electron microscopy (SEM) showed the
formation of crystals with dimensions around 2 microns and energy dispersive
X-ray (EDS) analysis indicated a nearly homogeneous composition, close to
the stoichiometric one. Field cooled magnetic measurement were held in a
SQUID magnetometer under a 10 kOe applied field, in order to sweep out
domain walls contribution to the magnetization curve.

Table \ref{tab1} displays the various compounds analyzed in the present
work, and the respective reference. Most of them were taken from the
literature. The choice was made such as to cover a wide range of $T_{c}$
values (within the ferromagnetic region, roughly between 100 and 400 K), but
was random in any other aspect. It is important to emphasize some
differences between the compounds, such as the $Mn^{3+}/Mn^{4+}$
concentration, distinct preparation processes, different ionic radii of
divalent ions, etc. Experimental details can be found in the list of
references.

The experimental results from the samples listed in table \ref{tab1} were
compared to the calculated total magnetization, $M_q$, given by:

\begin{equation}
M_{q}=(1-x)M_{q}^{3+}+xM_{q}^{4+}  \label{eq4}
\end{equation}
where $1-x$ and $x$ are, respectively, Mn$^{3+}$ and Mn$^{4+}$
concentrations. This equation, along Eqs. (\ref{eq2}) and (\ref{eq2a}) are
solved simultaneous and self-consistently in the mean-field approximation,
where the magnetic fields action on the 3+ and 4+ ions are given,
respectively, by $B^{3+}=B_{0}+\lambda ^{3+}M_{q}^{3+}$ and $%
B^{4+}=B_{0}+\lambda ^{4+}M_{q}^{4+}$, where $B_{0}$ is the external field.
The molecular field parameters $\lambda ^{3+}$ and $\lambda ^{4+}$ are, with
the value of $q$, input quantities which are varied to seek the best
agreement between the experimental and calculated results. It is also
important to mention that the analysis was carried on exactly the same data
as they appear in the original papers, and no normalization, re-arranging or
any kind of data treatment, such as smoothing, filtering, etc., took place.

Figure \ref{fig1} displays the experimental and calculated data for the
magnetization in La$_{0.89}$Sr$_{0.11}$Mn$_{0.93}$Cu$_{0.07}$O$_{3+\delta }$
compound. The various curves appearing correspond to different attempts of
using the Maxwell-Boltzmann, (i.e. q=1) statistics to reproduce experimental
data. These were: {\it approach 1 -} only Mn moments, without inter-lattice
interaction; {\it approach 2 - }Mn and Cu moment \cite{ref19}, with the same
intra-lattice interaction as in 1; {\it approach 3 - }Mn e Cu moment, with
inter- and intra-lattice interaction. The best agreement, however, is
obtained in {\it approach 4 - }only Mn moment, with the same interaction
parameters of approach 1, and $q=1.09$. Besides these attempts, we also
tried extensive models which consider the statistical distribution of Mn and
Cu ions in the crystal sites, but they all failed to explain the
experimental curve.

The above results clearly indicate a non-extensive behavior in these
compounds and, although the difference of the curves for $q\neq 1$ and $q=1$
is not too big, it served to us as a motivation to seek for further
examples. Therefore, we look up other $M$ vs. $T$ curves in the literature.
In figure \ref{fig2} (upper panel) we show a magnetic measurement, taken at
50 kOe, in La$_{0.875}$Sr$_{0.125}$MnO$_{3+\delta }$, reproduced from Ref. 
\cite{ref14a}, and the calculated $M_{q}$ curve for $q=0.86$ (superextensive
case). We also show the q=1 curve. In figure \ref{fig2} (lower panel) the $M$
vs. $T$ curve reproduced from Ref.\cite{ref14b} is shown for La$_{0.5}$Ba$%
_{0.5}$ MnO$_{3}$. The experiment was taken at 10 kOe. The best curve was
obtained for $q=1.07$ (subextensive case). Again, the $q=1$ case is shown
for comparison. The analysis of the other compounds shown in Table \ref{tab1}
follows in an analogous way.

In order to correlate the magnetic properties of manganites to the value of
the entropic parameter $q$ we show in Figure \ref{fig3} the experimental
value of $T_{c}$, plotted against $q$. The error in values of $q$ are less
than 2\%. Perhaps the most striking feature of this curve is its simplicity:
the data tend to fall on a straight line. This simplicity very much
contrasts with the usual complexity of physical behaviors observed in
manganites.

One can have a clue about the correlation between $T_{c}$ and $q$ shown in
Figure \ref{fig3} using a integral representation of the density matrix \cite
{integral} and making $B\rightarrow 0$.{\bf \ }For the case of a single
magnetic lattice, an expression for the $q$-dependent magnetic
susceptibility can be derived: $\chi _{q}=\partial M_{q}/\partial
B_{0}=C^{(q)}/(T^{\ast }-T_{c}^{(q)})$, where $T_{c}^{(q)}=T_{c}^{(1)}q=C%
\lambda q$\ and $C^{(q)}=C^{(1)}q$, revealing the linearity between $T_{c}$\
and $q,$\ still observed in our case.

In summary, we have found that the magnetic properties of manganites are
better described by the generalized statistics of Tsallis, instead the
traditional Maxwell-Boltzmann statistics. The entropic parameter $q$ is a
``measure'' of the degree of non-extensivity in the system, caused by
fractality and/or long-range interactions. These features have been invoked
in very recent works (references from \cite{ref17b} to \cite{ref17j}) and
recognized as essential aspects for the understanding of the magnetic and
transport properties of these materials, although no attempt has been made
to apply Tsallis formalism before the present work. At the moment, $q$ must
be regarded as a ``phenomenological'' parameter and its deviation from unit
is a confirmation of those features leading to non-extensivity in these
materials, and therefore in accordance to current models. In short, the use
of Tsallis statistics to analyze manganites can guide the development of new
models where the entropic parameter is identified to other fundamental
quantities of the problem.

The authors acknowledge the support from CAPES, CNPq and FAPERJ. We are
thankful to Professor C. Tsallis and Dr. V. Amaral for useful discussion.

\begin{figure}[tbp]
\caption{Experimental and calculated magnetization curves for La$_{0.89}$Sr$%
_{0.11}$Mn$_{0.93}$Cu$_{0.07}$O$_{3+\protect\delta }$. Dashed lines: {\em %
approach 1}; short dashed: {\em approach 2}; dash-doted: {\em approach 3}
and, solid line: Tsallis statistics result. See text for details.}
\label{fig1}
\end{figure}

\begin{figure}[tbp]
\caption{Experimental and calculated $(q=1\;{\rm and}\;q\neq 1)$
magnetization curves, for La$_{0.875}$Sr$_{0.125}$MnO$_{3+\protect\delta }$,
reprinted from \protect\cite{ref14a} (upper panel), and La$_{0.5}$Ba$_{0.5}$%
MnO$_{3}$, reprinted from \protect\cite{ref14b} (lower panel). The inset
displays the U$_{q}$ value [Eq. (2)], as a T$^{\ast }$ function. See text
for details.}
\label{fig2}
\end{figure}

\begin{figure}[tbp]
\caption{Correlation between $T_{c}$ and $q$. The straight line is only a
guide to the eyes.}
\label{fig3}
\end{figure}

\begin{table}[tbp]
\caption{Manganites compounds analyzed in the present work and the
respective references. The choice of compounds was made such as to cover a
wide range of $T_c$ values within the ferromagnetic phase, but it was random
in any other aspect.}
\label{tab1}
\begin{tabular}{|c|c|}
\hline
Compound & Reference \\ \hline\hline
La$_{0.7}$Sr$_{0.3}$Mn$_{0.9}$Ru$_{0.1}$O$_{3}$ & \cite{ref11} \\ \hline
La$_{0.5}$Ca$_{0.5}$MnO$_{3}$ & \cite{ref12} \\ \hline
La$_{0.83}$Sr$_{0.17}$Mn$_{0.98}$Fe$_{0.02}$O$_{3}$ & \cite{ref13} \\ \hline
La$_{0.62}$Y$_{0.07}$Ca$_{0.31}$MnO$_{3+\delta }$ & \cite{ref14} \\ \hline
La$_{0.875}$Sr$_{0.125}$MnO$_{3+\delta }$ & \cite{ref14a} \\ \hline
La$_{0.5}$Ba$_{0.5}$MnO$_{3}$ & \cite{ref14b} \\ \hline
La$_{0.75}$Ba$_{0.25}$MnO$_{3}$ & \cite{ref14b} \\ \hline
La$_{0.89}$Sr$_{0.11}$MnO$_{3+\delta }$ & Present work \\ \hline
La$_{0.89}$Sr$_{0.11}$Mn$_{0.93}$Cu$_{0.07}$O$_{3+\delta }$ & Present work
\\ \hline
\end{tabular}
\end{table}

\end{document}